\documentclass[prb,twocolumn,amsmath,amssymb, superscriptaddress]{revtex4}
\usepackage[pdftex]{graphicx}
 \usepackage{amsmath}
\usepackage[utf8x]{inputenc}
\usepackage[T1]{fontenc}
\usepackage{subfigure}
\usepackage{amssymb}
\usepackage{flushend}
 \usepackage{tikz}
\usetikzlibrary{decorations.markings}
\usepackage{amsfonts}
\usepackage{bm}
 \usepackage{amsmath} 
\usepackage{amsfonts} 
\usepackage{amssymb, mathrsfs}
\usepackage{braket}
\usepackage{graphicx} 

\usepackage{bbm}
\def\beq{\begin{equation}}
\def\eeq{\end{equation}}
\def\bsp{\begin{split}}
\def\esp{\end{split}}
\def\bea{\begin{eqnarray}}
\def\eea{\end{eqnarray}}
\def\ba{\begin{array}}
\def\ea{\end{array}}

\def\dg{\dagger}

\def\lb{\left(}
\def\rb{\right)}

\def\l.{\left.}
\def\r.{\right.}

\def\ra{\rangle}
\def\la{\langle}

\def\bo{\bold{k}}

\def\dg{\dagger}

\begin{document}
\author{S. A. Owerre}

\title{Topological Magnon Bands   in  Ferromagnetic Star Lattice}
\affiliation{ Perimeter Institute for Theoretical Physics, 31 Caroline St. N., Waterloo, Ontario N2L 2Y5, Canada.}
\email{solomon@aims.ac.za}
\affiliation{ African Institute for Mathematical Sciences, 6 Melrose Road, Muizenberg, Cape Town 7945, South Africa.}

\begin{abstract}
 The experimental observation of topological magnon bands and thermal Hall effect in a kagom\'e lattice ferromagnet Cu(1-3, bdc) has inspired the search for topological magnon effects in various insulating ferromagnets that lack an inversion center allowing a Dzyaloshinskii-Moriya (DM) spin-orbit interaction. The star lattice (also known as the decorated honeycomb lattice) ferromagnets is an ideal candidate for this purpose because it is a variant of the kagom\'e lattice with additional links that connect the up-pointing and down-pointing triangles. This gives rise to twice the unit cell of the kagom\'e lattice, hence a more interesting topological magnon effects. In particular,  the triangular bridges on the star lattice can be coupled either ferromagnetically or antiferromagnetically which is not possible on the kagome lattice ferromagnets. Here, we study DM-induced topological magnon bands, chiral edge modes, and thermal magnon Hall effect on the star lattice ferromagnet in different parameter regimes. The star lattice  can also be visualized as the parent material from which topological magnon bands can be realized for the kagom\'e and honeycomb lattices in some limiting cases.
\end{abstract}

\maketitle
\section{ Introduction}
Topological magnon matter is the magnonic analog of topological fermionic matter. In contrast, magnons are charge-neutral bosonic excitations  of ordered quantum magnets. Thus, the transport properties of magnons are believed to be the  new direction for dissipationless transports in insulating ferromagnets applicable to modern technology such as magnon spintronics and magnon thermal devices \cite{kru,kru1,kru2}. In insulating collinear quantum magnets the DM  interaction  \cite{dm,dm1} arising from spin-orbit coupling (SOC) \cite{dm1} is the key ingredient that leads to thermal magnon Hall effect \cite{alex0, alex1, alex1a, alex6} and topological magnon bands \cite{ alex2, alex7, alex6a,  hir,shin,shin1,zhh,alex4a, alex4, jh, sol,sol1,kkim}. The DM interaction is present in magnetic systems that lack inversion symmetry between magnetic ions. The kagom\'e lattice is built with this structure, because  the midpoint of the bonds connecting two nearest-neighbour magnetic ions is not a center of inversion. Therefore, a DM interaction is intrinsic to the kagom\'e lattice.  The thermal magnon Hall effect induced by the DM interaction was first observed experimentally in a number of three-dimensional (3D) ferromagnetic pyrochlores  --- Lu$_2$V$_2$O$_7$, Ho$_2$V$_2$O$_7$, and In$_2$Mn$_2$O$_7$ \cite{alex1,alex1a}.  Subsequently, the  thermal magnon Hall effect was realized in a  kagom\'e ferromagnet Cu(1-3, bdc) \cite{alex6} followed by the first experimental realization of topological magnon bands \cite{alex6a} in the same material.  On the other hand, no topologcal magnon bands have been observed in pyrochlore ferromagnets, but recent studies have proposed Weyl magnons  in  3D pyrochlore ferromagnets \cite{mok,su}.

 In the same way a pyrochlore lattice can be visualized  as an alternating kagom\'e and triangular layers, the star lattice (also known as the decorated honeycomb lattice) can be  visualized as an interpolating lattice between the honeycomb lattice and the kagom\'e lattice. The number of sites per unit cell in star lattice is six, three times as in honeycomb and twice as in kagome. Most importantly, the star lattice plays a prominent role in different branches of physics.  The Kitaev model on the star lattice has an exact solution as a chiral spin liquid \cite{zheng1}. The star lattice also plays an important role in ultracold atoms \cite{zheng6}. In particular, quantum magnets \cite{zheng,zheng1, zheng3, zheng4, zheng6} and topological insulators \cite{ zheng2,zheng5, zheng5a} on the star-lattice show distinctive remarkable features different from  the kagome and honeycomb lattices. 
 
 Motivated by these interesting features on the star lattice, the study of topological  magnon bands  on the star lattice is necessary and unavoidable.  In this work, we study the  topological magnon bands and the thermal magnon Hall response in insulating quantum  ferromagnet on the star lattice.   We show that lack of an inversion center allows a  DM interaction  between the midpoint of the triangular bonds. The DM interaction induces a fictitious magnetic flux in the unit cell and leads to nontrivial topological magnon bands. This directly leads  to the existence of nonzero Berry curvatures and Chern numbers  accompanied by topologically protected gapless edge modes.  Indeed, it is feasible to synthesize   magnetic materials with the star structure and  directly confirm the present theoretical results.

\begin{figure}
\centering
\includegraphics[width=1\linewidth]{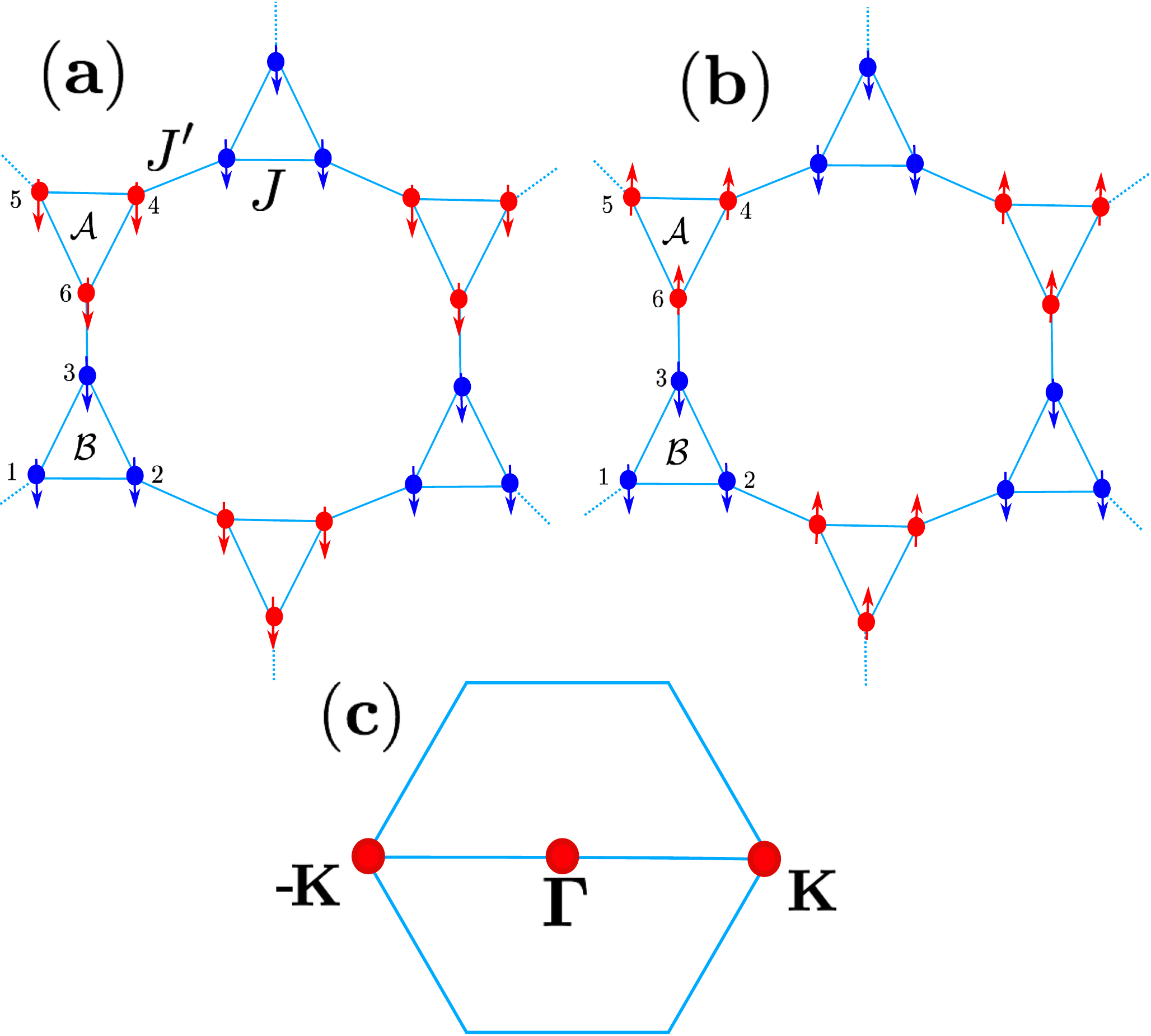}
\caption{Color online. Schematics of star lattice. (a) Coupled ferromagnets within and between triangles of the star lattice. (b) Coupled antiferromagnets between  the triangles. (c). The Brillouin zone of the star lattice. }
\label{fig1}
\end{figure} 
\begin{figure}
\centering
\includegraphics[width=.75\linewidth]{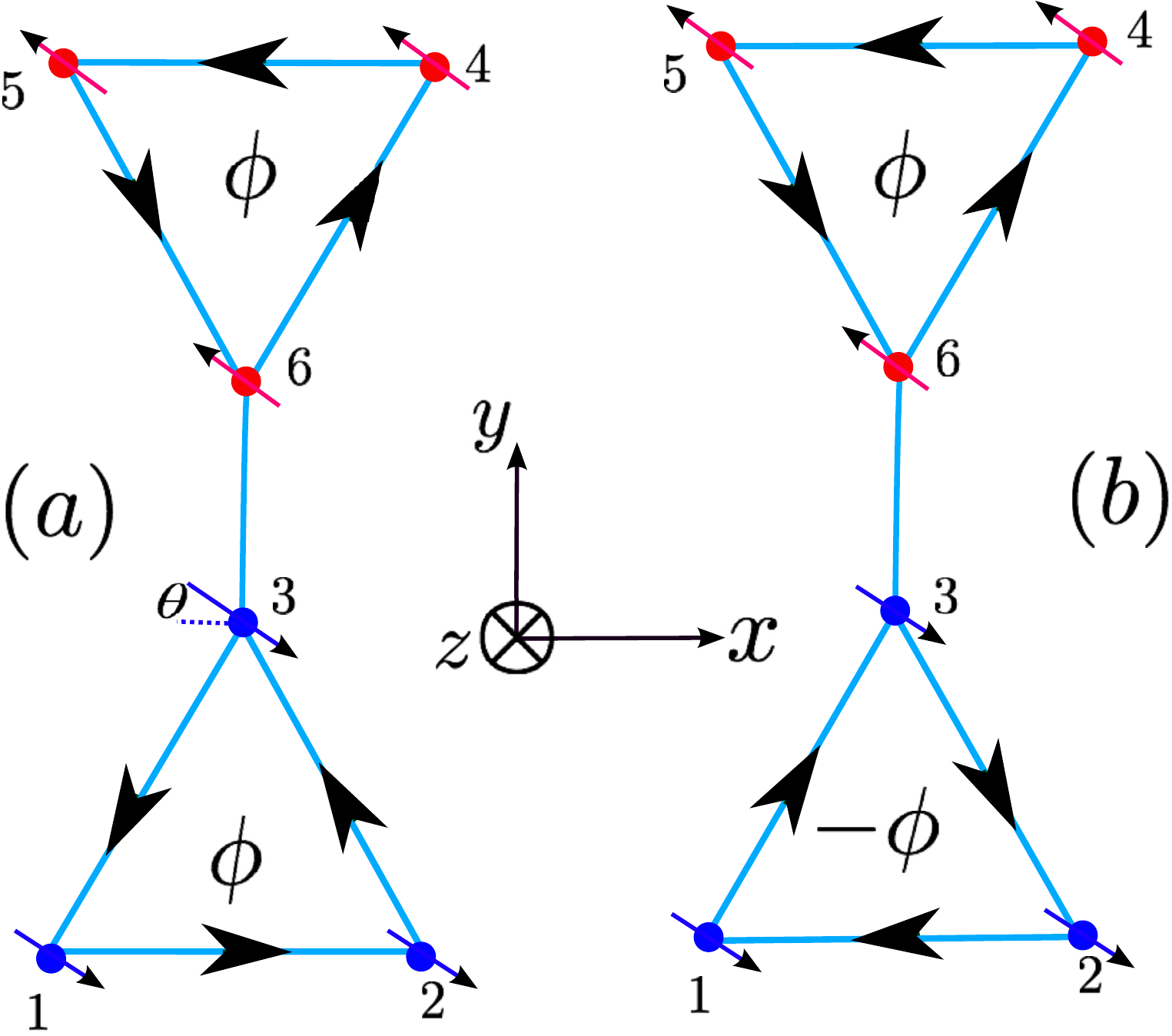}
\caption{The possible configurations of the DM-induced flux $\phi_{ij}$ (see Appendix \ref{free}) in the presence of a magnetic field, where $\theta$ is the field-induced canting angle. Bold arrows indicate the sign of the fictitious magnetic flux $\phi_{ij}$ and the small arrows show the magnetic-field-induced spin canting in the $x$-$z$ plane.  $(a)$ Uniform flux. $(b)$ Staggered flux.}
\label{fig2}
\end{figure}
\section{Ferromagnetic Hamiltonian}
Quantum magnets on the   star lattice are known to possess magnetic long-range  orders \cite{zheng, zheng4}. However, the effects of the intrinsic DM perturbation have not been  considered on the star lattice. Here, we study a ferromagnetic model which is a  magnetically ordered state on the star lattice. The Hamiltonian is given by
\begin{align}
\mathcal H&=\mathcal H_0+\mathcal H_{Z}+ \mathcal H_{\text{pert}},
\label{h}
\end{align}
where
\begin{align}
\mathcal H_0=-J\sum_{\la ij\ra}{\bf S}_{i}\cdot{\bf S}_{j}+J^\prime\sum_{\la ij\ra}{\bf S}_{i}\cdot{\bf S}_{j},
\end{align}
and  $J,J^\prime$ are exchange couplings within sites on the triangles ``$\Delta$''  and between triangles ``$\Delta \leftrightarrow \nabla$'' as shown in Fig.~\ref{fig1}. The Zeeman magnetic field is given by
\begin{align}
\mathcal H_{Z}=- \bold{H}\cdot\sum_{i} {\bf S}_i,
\end{align}
where $\bold{H}=h\bold{\hat z}$ with $h=g\mu_B H$ is the strength of the out-of-plane magnetic field. The last term $\mathcal H_{\text{pert}}$ represents all the perturbative interactions to the Heisenberg exchange. The DM interaction is usually the dominant perturbative anisotropy. It is allowed on the star-lattice due to lack of inversion center between magnetic ions according to the Moriya rule  \cite{dm1}.  The magneto-crystalline anisotropy is second order in perturbation which can be neglected.  Therefore, we will consider only the DM perturbation term. Hence,
\begin{align}
\mathcal H_{\text{pert}}= \sum_{ \la ij\ra}\bold D_{ij} \cdot{\bf S}_{i}\times{\bf S}_{j}.
\end{align}
 There are two ferromagnetic ordered states on the star-lattice. The first one corresponds to  $J>0$ and $J^\prime<0$, i.e. fully polarized ferromagnets on sublattice $\mathcal A$ (down pointing triangles indicated with blue sites) and $\mathcal B$ (up pointing triangles indicated  with red sites) as shown in Fig.~\ref{fig1}.  The second one corresponds to  $J>0$ and $J^\prime>0$, i.e., antiferromagnetic interaction between triangles and ferromagnetic interaction on each triangle. In the latter case the spins on sublattice $\mathcal A$ are oriented  in the opposite direction to those on sublattice $\mathcal B$, and they also cant along the magnetic field direction as shown in Fig.~\ref{fig2}. As we will show in the subsequent sections, the latter case recovers the former case at the saturation field $h_s=2J^\prime S$.
\section{Results}
\subsection{Ground state energy}
 At zero field the spins  are  aligned along the star plane chosen as the $x$-$y$ plane with the quantization axis chosen along the $x$-direction. The ground state is a collinear ferromagnet unaffected by the DM interaction. However, if the sublattices $\mathcal A$ and $ \mathcal B$ are coupled antiferromagnetically, that is $J>0$ and $J^\prime>0$, then a small magnetic field induces canting along the direction of the field and the ground state is no longer the collinear ferromagnets. The collinear ferromagnet is only recovered at the  saturation field $h_s$. In the large-$S$ limit, the spin operators  can be approximated as classical vectors,  written as
 $\bold{S}_{\tau}= S\bold{n}_\tau$, where $\bold{n}_\tau=\lb\sin\theta\cos\vartheta_\tau, \sin\theta\sin\vartheta_\tau,\cos\theta \rb$
 is a unit vector and $\tau$ denotes the down ($\mathcal A$)  and up ($\mathcal B$) triangles, with $\vartheta_{\mathcal A}=0$ and $\vartheta_{\mathcal B}=\pi$ and $\theta$ is the magnetic-field-induced canting angle.   As the system is ordered ferromagnetically on each triangle,  the DM interaction does not contribute to the classical energy  given by
\begin{align}
e_0=-JS^2+\frac{J^\prime S^2}{2}\cos2\theta- hS\cos\theta,
\end{align}
 where $e_0=E/6N$ is energy per site and $N$ is the number of sites per unit cell. Minimizing the classical energy yields the canting angle $\cos\theta= h/h_s$.
 
 \subsection{Magnetic excitations}
 In the low temperature regime the magnetic excitations  above the classical ground state are magnons and higher order magnon-magnon interactions are negligible. Thus, the linearized Holstein-Primakoff (HP) spin-boson representation \cite{HP} is valid. Due to the magnetic-field-induced spin canting, we have to rotate the coordinate axes such that the $z$-axis coincides with the local direction of the classical polarization. This involves rotating  the spins from laboratory frame to local frame  by the spin oriented angles $\vartheta_\tau$  about the $z$-axis. This rotation is followed by  another rotation about the $y$-axis with the canting angle $\theta$, and the resulting transformation is given by   
\begin{align}
&S_{i\tau}^x=\pm S_{i\tau}^{\prime x}\cos\theta \pm S_{i\tau}^{\prime z}\sin\theta,\label{trans}\nonumber\\&
S_{i\tau}^y=\pm S_{i\tau}^{\prime y},\\&\nonumber
S_{i\tau}^z=- S_{i\tau}^{\prime x}\sin\theta + S_{i\tau}^{\prime z}\cos\theta,
\end{align}
where $+(-)$  applies to the spins residing on the sublattice $\mathcal A$ ($\mathcal B$) triangles respectively. It is easily checked that this rotation does not affect the collinear ferromagnetic ordering on each triangle, that is the $J$ term. 

Usually, in collinear ferromagnets the perpendicular-to-field DM component does to contribute to the free magnon theory because  the magnetic moments are polarized along the  magnetic field direction \cite{alex1, alex1a,alex6}. Therefore, only the parallel-to-field DM component has a significant contribution to the free magnon dispersion \cite{ alex2, alex7, alex6a,  hir,shin,shin1,zhh,alex4a, alex4, jh, sol,sol1,kkim,alex0, alex1, alex1a, alex6}. In the present model, however, the situation is different. Due to antiferromagnetic coupling between triangles an applied magnetic field induces two spin components --- one parallel to the field and the other perpendicular to the field. As we pointed out above, the DM interaction has a finite contribution to the free magnon model only when the magnetic moments are along its direction.  Therefore the  parallel-to-field DM component (${\bf D}\parallel {\bf H}$)  and the perpendicular-to-field DM component (${\bf D}\perp {\bold H}$) will have a finite contribution to the free magnon theory due to spin  canting. The former is  rescaled as  $D_{\parallel,\theta}\to D\cos\theta$ and the latter is rescaled as $D_{\perp,\theta}\to D\sin\theta$ by the magnetic field. Implementing the spin transformation \eqref{trans} the terms that contribute to the free magnon model are given by \begin{align}
&\mathcal H_{0}=-J\sum_{\la i, j\ra\tau}{\bf S}_{i\tau}^\prime\cdot{\bf S}_{j\tau}^\prime
 \nonumber\\&-J^\prime\sum_{\la i,j\ra\tau\tau^\prime}[\cos 2\theta (S_{i\tau}^{\prime x}S_{j\tau^\prime}^{\prime x}-S_{i\tau}^{\prime z}S_{j\tau^\prime}^{\prime z})+S_{i\tau}^{\prime y}S_{j\tau^\prime}^{\prime y}],\\& ~\mathcal H_Z= -h_{\theta}\sum_{i\tau} S_{i\tau}^{\prime z},\\&
\mathcal H_{\text{pert},\parallel}= D_{\parallel,\theta}\sum_{\la i, j\ra\tau}\chi_{ij\tau}^{\prime z},~\mathcal H_{\text{pert},\perp}=  \pm D_{\perp,\theta}\sum_{\la i, j\ra\tau}\chi_{ij\tau}^{\prime z},
\end{align}
where  $h_{\theta}=h\cos\theta$ and $\chi_{ij\tau}^{\prime z}$ is the $z$-component of the vector spin chirality $[{\bf S}_i^\prime\times {\bf S}_j^\prime]_z$.  We note that  $H_{\text{pert},\parallel}$  and  $H_{\text{pert},\perp}$ DM components are oriented perpendicular to the bond due to the  magnetic field. 
From these expressions it is evident that at zero magnetic field ($\theta=\pi/2$) the spins are along the $x$-$y$ plane of the star lattice, therefore only the DM component parallel to the spin axis  has a finite contribution to the free magnon model. The same is true  at the saturation field ($\theta=0$) when the spins are fully aligned along the $z$-direction. In Appendix \ref{free} we have shown the tight binding magnon model.  In the following section we will consider the two field-induced spin components separately.

  \begin{figure}
\centering
\includegraphics[width=1\linewidth]{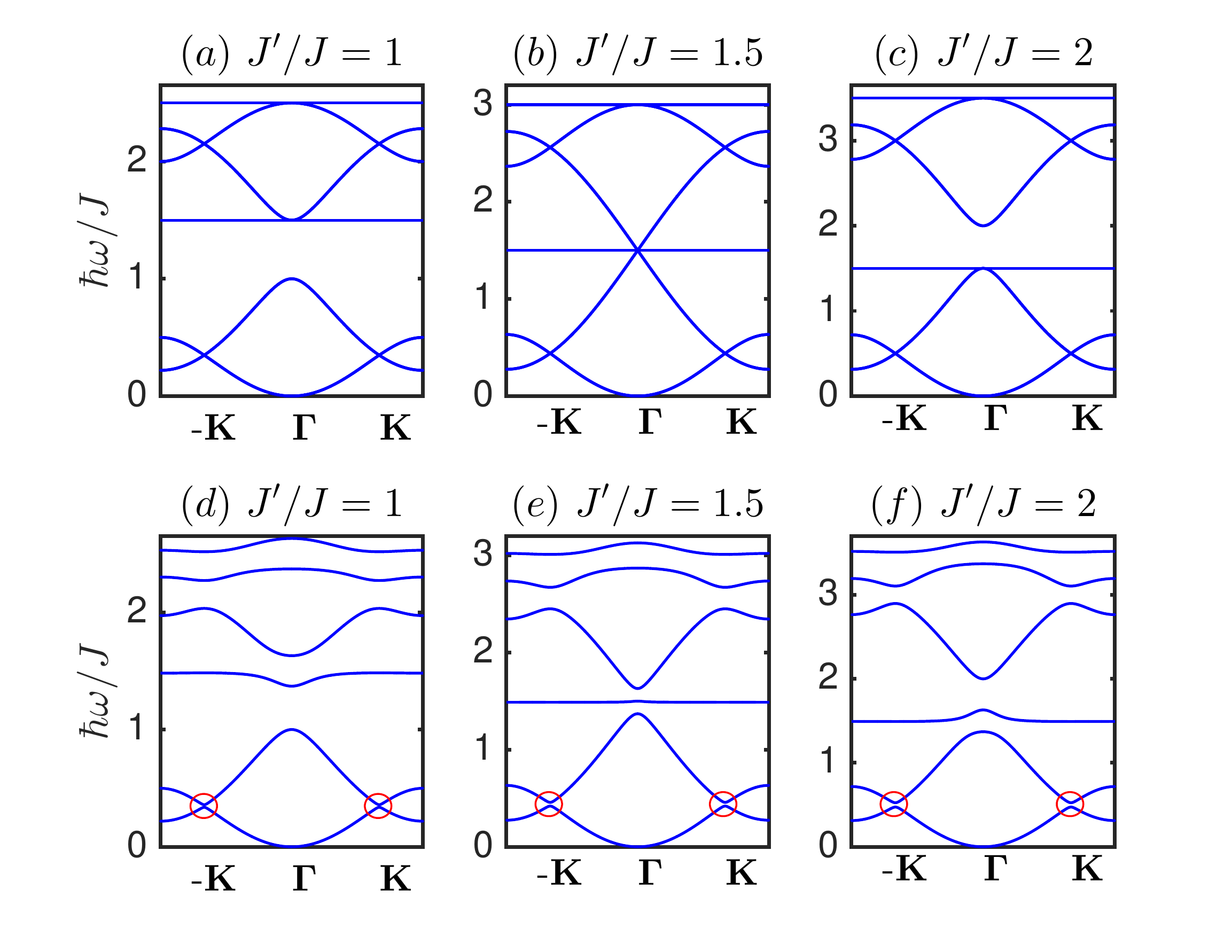}
\caption{Color online. Magnon bands in the fully polarized ferromagnet at the saturation field  $h=h_s$.  Top panel: $ D=0$. Bottom panel: $D/J=0.15$. The circled points are gapped. }
\label{bandx1}
\end{figure}

\begin{figure}
\centering
\includegraphics[width=1\linewidth]{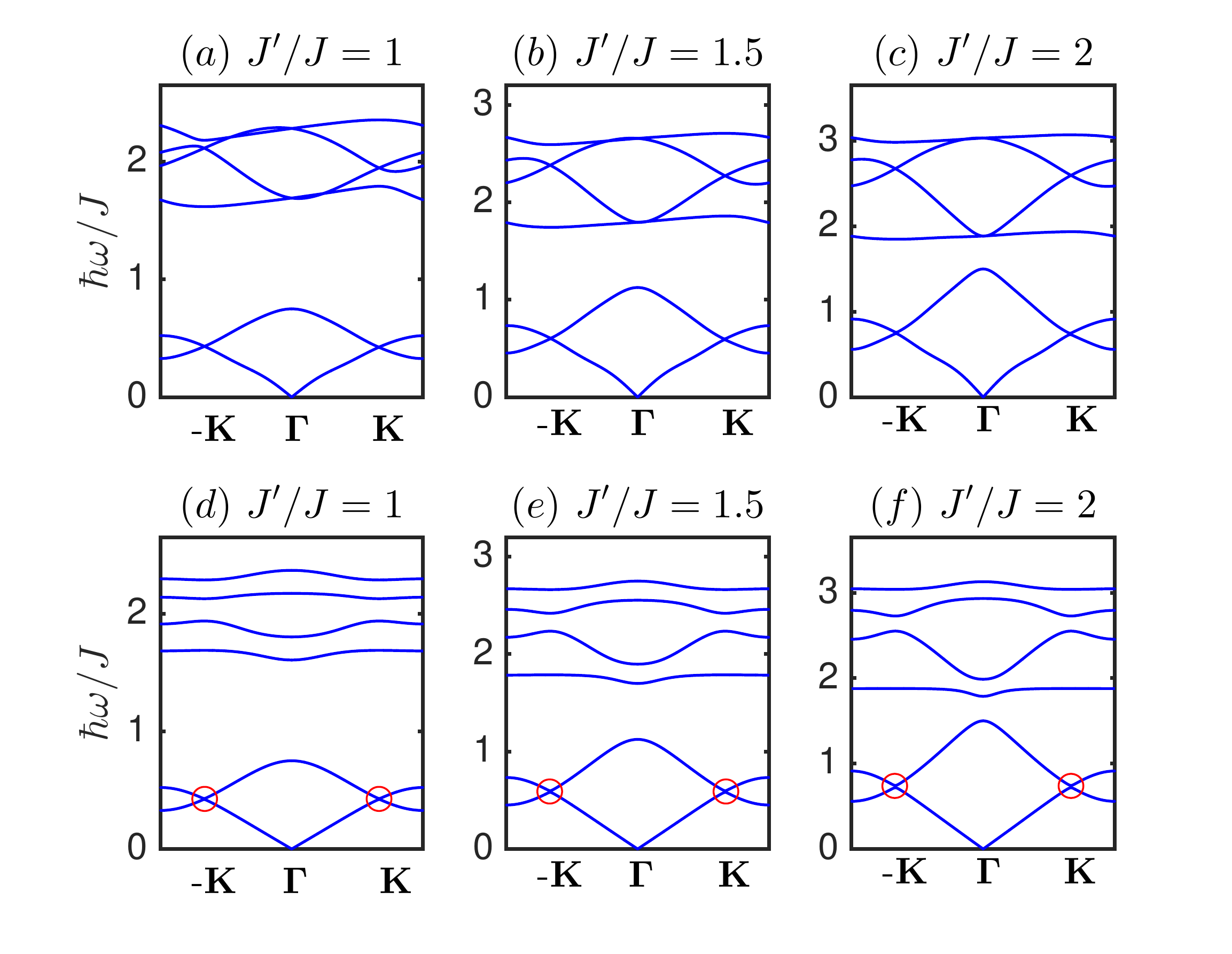}
\caption{Color online. Magnon bands in the canted antiferromagnet for $h<h_s=0.75J$. Top panel: Perpendicular-to-field  spin components. Bottom panel: Parallel-to-field  spin components. The DM value is $D/J=0.15$. The circled points are gapped.}
\label{bandz1}
\end{figure}

 \subsection{Topological magnon bands}
We note that the first topological magnon bands have been measured in the kagom\'e ferromagnet Cu(1-3, bdc)  \cite{alex6a}, thus paving the way to search for topological magnon bands in other systems. 
In the following we study the topological magnon bands in the star lattice ferromagnet. We  consider spin-$1/2$ and take $J$ as the unit of energy  while varying $J^\prime/J$.  We also take the DM value of the kagom\'e ferromagnet Cu(1-3, bdc) $D/J=0.15$ \cite{alex6,alex6a}. In Fig.~\ref{bandx1} we have shown the magnon bands at  $D=0$ and $D/J=0.15$ with varying $J^\prime/J$.   In the former (top panel) the lower bands for $J^\prime/J<1.5$ in Fig.~\ref{bandx1}(a) looks like Dirac magnon on the honeycomb lattice \cite{mag}. On the other hand, for $J^\prime/J>1.5$  the bands in Fig.~\ref{bandx1}(c) resemble two copies of magnon bands on the kagom\'e lattice ferromagnet with a flat band and two dispersive Dirac magnon bands on each copy. The gap closes at $J^\prime/J=1.5$
as shown in Fig.~\ref{bandx1}(b). For nonzero DM interaction (lower panel) the the magnon bands are separated by a finite energy gap proportional to the DM interaction  in all the parameter regimes. Notice that the collinear ferromagnet at $h=h_s$ has a quadratic dispersion (Goldstone mode) at the ${\bf \Gamma}$ point due spontaneous breaking of U(1) symmetry about the $z$-axis.  In contrast, for $h<h_s$ the spins are canted due to antiferromagnetic interaction between the triangles. This leads to two spin components ---  perpendicular-to-field  spin components and  parallel-to-field  spin components. In Fig. \ref{bandz1} we have shown the magnon bands in the canted antiferromagnetic phase. Indeed, we recover a linear Goldstone mode at the ${\bf \Gamma}$ point which signifies an antiferromagnetic spin order.  In this case the model is no longer an analog of topological fermion insulator on the star lattice \cite{zheng2, zheng5, zheng5a}.  Furthermore, the  perpendicular-to-field  spin components (top panel of Fig. \ref{bandz1}) show gapless magnon bands even in the presence of DM interaction.  A similar gapless magnon bands was reported in the kagom\'e ferromagnet Cu(1-3, bdc) \cite{alex6a} when the spins are polarized  along the kagom\'e plane by an in-plane magnetic field. In contrast, the parallel-to-field spin components (bottom panel of Fig. \ref{bandz1}) show a finite gap separating the magnon bands due to the presence of DM interaction along the spin polarization.
\subsection{Chiral edge modes}
\begin{figure}
\centering
\includegraphics[width=1\linewidth]{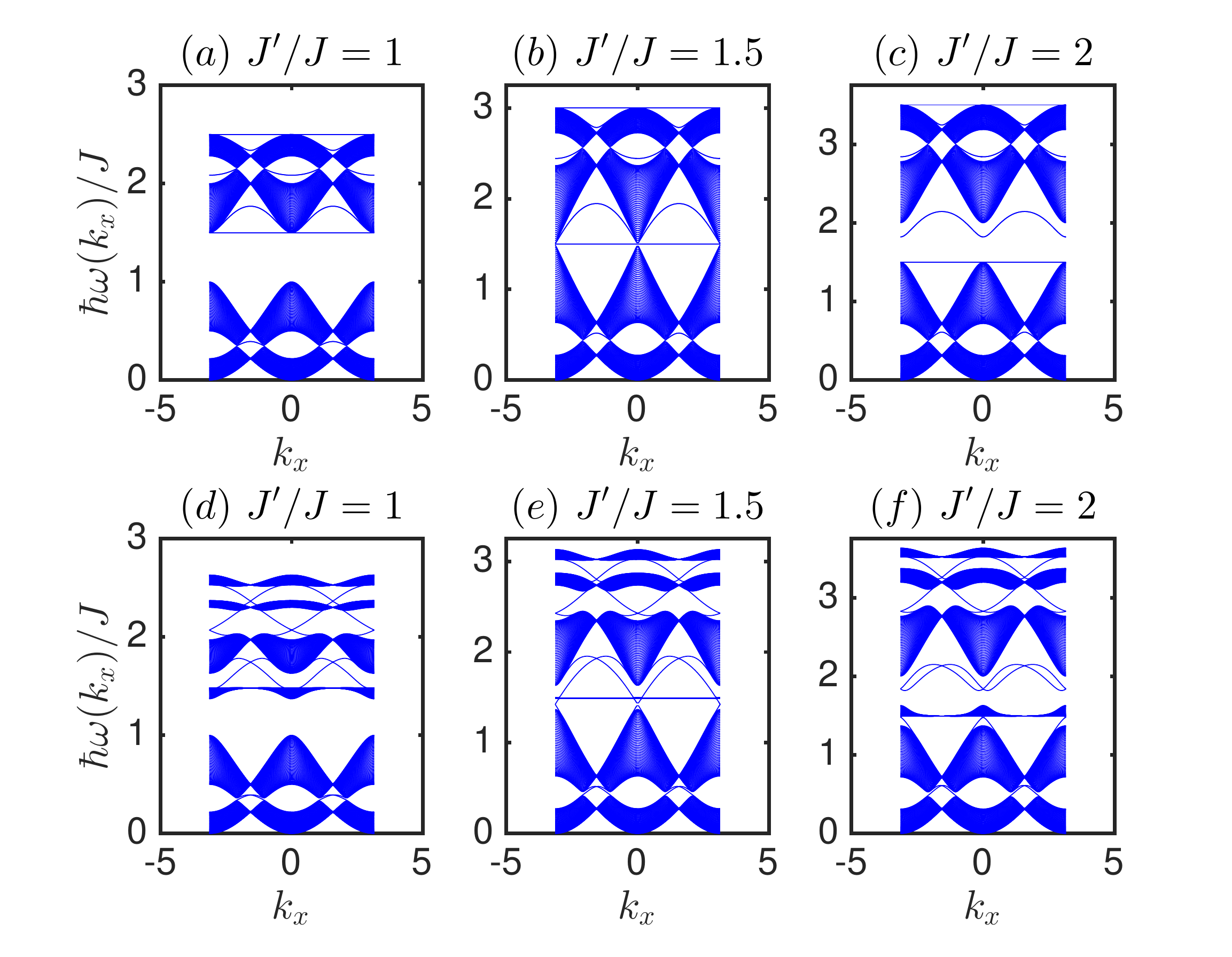}
\caption{Color online. The corresponding chiral magnon edge modes for Fig.~\ref{bandx1}.}
\label{eg1}
\end{figure}

\begin{figure}
\centering
\includegraphics[width=1\linewidth]{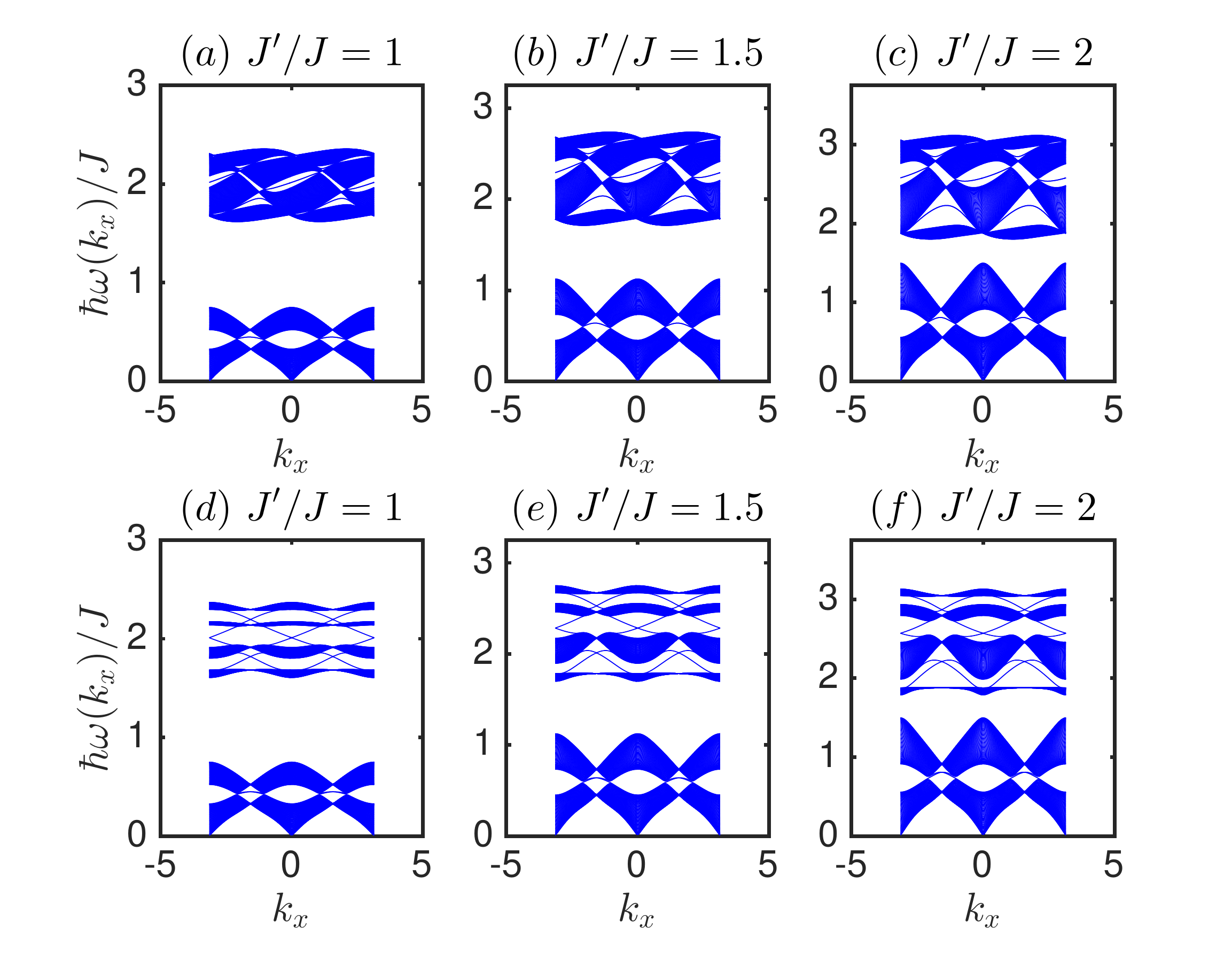}
\caption{Color online. The corresponding chiral magnon edge modes for Fig.~\ref{bandz1}.}
\label{eg2}
\end{figure}

The topological aspects of gap Dirac points can be studied  by the defining the Berry curvature and the Chern number. In the present model we started from an antiferromagnetic coupled ferromagnets and then recovered a collinear ferromagnetic model. Therefore the Hamiltonian has an off-diagonal terms and the diagonalization requires the generalized Bogoliubov transformation and the eigenfunctions define the Berry curvatures and Chern numbers (see Appendix \ref{berry}). 
The Chern number vanishes for all bands  at zero DM interaction (top panel of Fig. \ref{bandx1}).  We also find that the Chern number vanishes for the magnon bands of perpendicular-to-field spin components (top panel of Fig. \ref{bandz1}). This signifies a trivial magnon insulator.  However, we find  nonzero Chern numbers  for the other magnon bands (bottom panel of Figs. \ref{bandx1} and  \ref{bandz1}) which defines a topological magnon insulator. A nonzero Chern number is associated with  topological chiral gapless magnon edge modes which appear at the DM-induced gaps as shown in Figs. \ref{eg1} and \ref{eg2}. The edge modes are solved  for a strip geometry on the star-lattice with open boundary conditions along the $y$ direction and infinite along the $x$ direction.

 \begin{figure}
\centering
\includegraphics[width=1\linewidth]{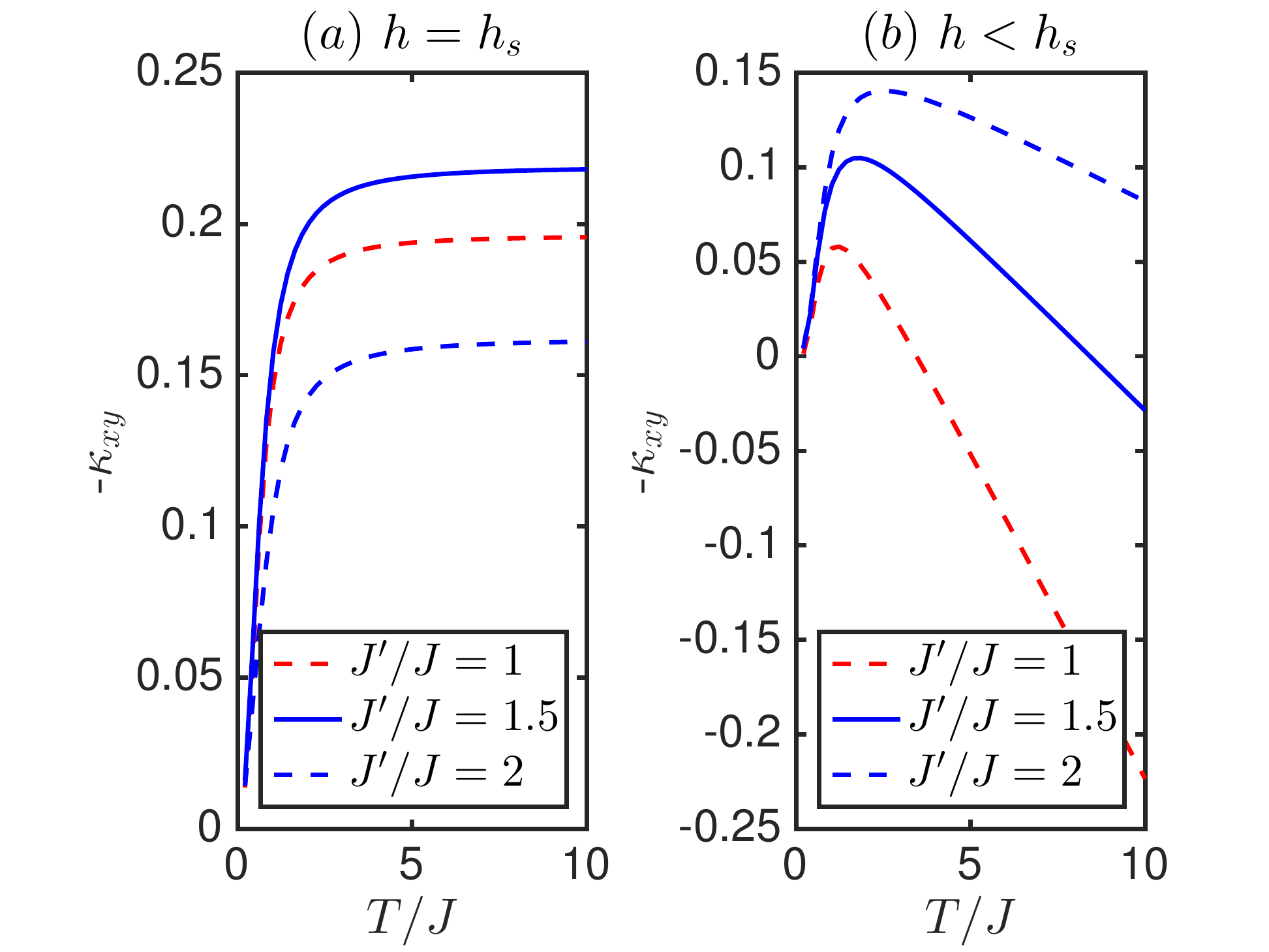}
\caption{Color online. Low temperature dependence of the thermal Hall conductivity for several values of $J^\prime/J$ and $D/J=0.15$. }
\label{hall}
\end{figure}

\subsection{Thermal magnon Hall effect}
Inelastic neutron scattering experiment has measured the thermal Hall conductivity in the kagom\'e and pyrochlore ferromagnets \cite{alex1,alex1a, alex6}. Theoretically, the thermal Hall effect in insulating ferromagnets is   understood as a consequence of the topological magnon bands induced by the DM interaction \cite{alex0}. A temperature gradient $-\boldsymbol\nabla T$ induces a transverse heat current $\bold J^Q$ and the DM-induced Berry curvature acts as an effective magnetic field that deflects the propagation of magnon in the system giving rise to a thermal Hall effect similar to Hall effect in electronic systems.    From linear response theory, one obtains $\mathcal J_{\alpha}^Q=-\sum_{\beta}\kappa_{\alpha\beta}\nabla_{\beta} T$, where $\kappa_{\alpha\beta}$ is the thermal conductivity  and the transverse component $\kappa_{xy}$  is associated with thermal Hall conductivity given explicitly as \cite{alex2,shin1}
\begin{align}
\kappa_{xy}=-\frac{k_B^2 T}{\hbar V}\sum_{\bo}\sum_{\alpha=1}^N \lb c_2[ g\lb \omega_{\bo \alpha}\rb]-\frac{\pi^2}{3}\rb\Omega_{xy;\alpha}(\bold k), 
\label{thm}
\end{align}
where $V$ is the volume of the system, $k_B$ is the Boltzmann constant, $T$ is the temperature, $g(\omega_{\alpha\bo})=[e^{{\omega_{\alpha\bo}}/k_BT}-1]^{-1}$ is the Bose function, and $c_2(x)$ is defined as  \bea c_2(x)=(1+x)\lb \ln \frac{1+x}{x}\rb^2-(\ln x)^2-2\text{Li}_2(-x),\eea with $\text{Li}_2(x)$ being the  dilogarithm. The Berry curvature $\Omega_{xy;\alpha}(\bold k)$ is defined in Appendix \ref{berry}.  The thermal Hall conductivity is finite only in the fully polarized  collinear ferromagnets at $h=h_s$ and the parallel-to-field spin components in the canted antiferromagnet for $h<h_s$. As shown in Fig.~\ref{hall} the thermal Hall conductivity shows a sharp peak and a sign change in the canted phase and vanishes at zero temperature as there are no thermal excitations.  This is consistent with the trend observed in previous experiments on the kagom\'e and pyrochlore ferromagnets \cite{alex1, alex1a,alex6}.

\section{Conclusion} 
The star lattice has attracted considerable attention in recent years as an exact solution of Kitaev model \cite{zheng1}. Chiral spin liquids, topological fermion insulators and quantum anomalous Hall effect have been proposed \cite{zheng,zheng1, zheng3, zheng4, zheng6,zheng2,zheng5, zheng5a}. However, there is no experimental realizations at the moment. In this work, we have contributed to the list of interesting proposals on the star lattice. We have shown that insulating quantum  ferromagnets on the star lattice  are candidates for   topological magnon insulators and thermal magnon Hall transports. We showed that the intrinsic DM interaction which is allowed on the star lattice gives rise to  magnetic excitations that exhibit nontrivial magnon bands with non-vanishing Berry curvatures and Chern numbers. We believe that the  synthesis of magnetic  materials with a star structure is feasible.   In fact,  experiment has previously realized polymeric iron (III) acetate as a star lattice antiferromagnet with both spin frustration and magnetic long-range order \cite{zheng}. A strong applied magnetic field is sufficient to induce a ferromagnetic ordered phase in this material and the topological magnon bands can be realized. Unfortunately,  inelastic neutron scattering is a bulk sensitive method and  the chiral magnon edge modes have not been measured in any topological magnon insulator \cite{alex6a}. It is possible that   edge sensitive methods such as  light \cite{luuk} or electronic \cite{kha} scattering method can see the chiral magnon edge modes in topological magnon insulators.
\section*{ Acknowledgments}
Research at Perimeter Institute is supported by the Government of Canada through Industry Canada and by the Province of Ontario through the Ministry of Research
and Innovation.
 \appendix
 \section{Free magnon theory}
 \label{free}
The corresponding free magnon model is achieved by mapping the spin operators  to boson operators \cite{HP}: 
$S_{i,\tau}^{\prime x}=\sqrt{S/2}(b_{i,\tau}^\dg+b_{i,\tau})$, $S_{i,\tau}^{\prime y}=i\sqrt{S/2}(b_{i,\tau}^\dg-b_{i,\tau})$, and $S_{i,\tau}^{\prime z}=S-b_{i,\tau}^\dg b_{i,\tau}$.   
The resulting free magnon  model is given by
\begin{align}
\mathcal H&=v_0\sum_{j,\tau}n_{j\tau} - v_t\sum_{\la ij\ra,\tau}(e^{-i\sigma \phi_{ij}} b^\dagger_{i\tau} b_{j\tau}+h.c.)\nonumber\\& -v^\prime\sum_{\la i j\ra\tau\tau^\prime}[( b_{i\tau}^\dagger b_{j\tau^\prime}+ h.c.)\cos^2\theta-( b_{i\tau}^\dagger b_{j\tau^\prime}^\dagger+ h.c.)\sin^2\theta],
\end{align}
where $n_{j\tau}=b_{j\tau}^\dagger b_{j\tau}$, $v_0=2v_s-v_s^\prime\cos 2\theta + h\cos\theta=2v_s +v_s^\prime$, $v_s(v_s^\prime)= JS(J^\prime S)$, ~$v_t=\sqrt{v_s^{ 2} +v_{D_{z(x),\theta}}^2}=JS/\cos(\phi_{ij}),~v_{D_{\parallel(\perp),\theta}}=D_{\parallel(\perp),\theta}S$;  $\phi_{ij}=\pm\phi=\arctan(D_{\parallel(\perp),\theta}/J)$ is a magnetic flux generated by the DM interaction within the triangular plaquettes, similar to Haldane model \cite{fdm}.  For ${\bf D}\parallel {\bf H}$, $\sigma=1$ and for ${\bf D}\perp {\bf H}$, $\sigma=\pm 1$ for sublattice $\mathcal A$ and $\mathcal B$ respectively. The configurations of $\phi_{ij}$ for both cases are depicted in Fig.~\ref{fig2}. The total flux in the dodecagon consisting of twelve sites is $-2\phi$ and $0$ respectively. Indeed,  $\phi_{ij}$ vanishes along the $J^\prime$ link as it contains no triangular plaquettes.  

The momentum space Hamiltonian can be written as $ \mathcal H=\frac{1}{2}\sum_{\bold k}\psi^\dagger_{\bold k}\cdot \boldsymbol{\mathcal{H}}(\bold k)\cdot\psi_{\bold k},$ with $\psi^\dagger_{\bold k}= (b_{\mu,\bold{k}}^{\dagger},\thinspace b_{\mu^\prime,\bold{k}}^{\dagger}, b_{\mu,-\bold{k}},\thinspace b_{\mu^\prime,-\bold{k}})$, where $\mu=1,2,3$ and  $\mu^\prime=4,5,6$. The Bogoliubov Hamiltonian  $\mathcal{H}(\bold k)$ is a $2N\times 2N$ matrix given by
\begin{align}
\boldsymbol{\mathcal{H}}(\bold k)= 
\begin{pmatrix}
\bold{A}(\bo, \phi) & \bold{B}(\bo)\\
\bold{B}^*(-\bo)&\bold{A}^*(-\bo,\phi)
\end{pmatrix},
\label{ham}
\end{align}
where  \begin{align}
\bold{A}(\bold k)= 
\begin{pmatrix}
\bold{a}(\sigma\phi) & \bold{b}_1(\bo)\\
\bold{b}_1(-\bo)&\bold{a}(\sigma\phi)
\end{pmatrix},~
\bold{B}(\bold k)= 
\begin{pmatrix}
{\bf 0} & \bold{b}_2(\bo)\\
\bold{b}_2(-\bo)& {\bf 0}
\end{pmatrix},
\end{align}

\begin{align}
&\bold{a}(\sigma\phi)= 
\begin{pmatrix}
v_0&-v_t e^{-i\sigma\phi}&-v_t e^{i\sigma\phi}\\
-v_t e^{i\sigma\phi}&v_0&-v_t e^{-i\sigma\phi}\\
-v_t e^{-i\sigma\phi}&-v_t e^{i\sigma\phi}&v_0
\end{pmatrix},\\&
\bold{b}_1(\bo)=-v_s^\prime\cos^2\theta 
\begin{pmatrix}
 e^{ik_2}&0&0\\
0& e^{ik_1}&0\\
0&0&1
\end{pmatrix},\\&
\bold{b}_2(\bo)=v_s^\prime\sin^2\theta 
\begin{pmatrix}
e^{ik_2}&0&0\\
0& e^{ik_1}&0\\
0&0&1\\
\end{pmatrix},
\end{align}
where $k_{1}=\bold{k}\cdot \bold{a}_{1}$ and $k_{2}=\bold{k}\cdot \bold{a}_{2}$. The lattice basis vectors are chosen as $\bold a_1=2\hat{\bold x}$ and  $\bold a_2= \hat{\bold x} + \sqrt{3}\hat{\bold y}$. At the saturation field $h=h_s (\theta=0)$ the spins are fully aligned along the $z$-axis  and  $\bold{b}_2(\bo)=0$.  We therefore recover  collinear ferromagnet along the  $z$-direction. The Hamiltonian is diagonalized below. 

\section{Berry curvature and Chern number}
\label{berry}
To diagonalize the Hamiltonian  we make a linear  transformation 
$\psi_\bo= \mathcal{P}_\bo Q_\bo$, 
where $\mathcal{P}_\bo$ is a $2N\times 2N$ paraunitary matrix and  $Q^\dg_\bo= (\mathcal{Q}_\bo^\dg,\thinspace \mathcal{Q}_{-\bo})$ with $ \mathcal{Q}_\bo^\dg=(\gamma_{\bo\mu}^{\dg},\thinspace \gamma_{\bo\mu^\prime}^{\dg})$ being the quasiparticle operators. The matrix $\mathcal{P}_\bo$ satisfies the relations,
\begin{align}
&\mathcal{P}_\bo^\dg \boldsymbol{\mathcal{H}}(\bo) \mathcal{P}_\bo=\mathcal{E}_\bo,\label{eqn1}\\ &\mathcal{P}_\bo^\dg \boldsymbol{\tau}_3 \mathcal{P}_\bo= \boldsymbol{\tau}_3,
\label{eqna}
\end{align}
where $\mathcal{E}_\bo= \text{diag}(\omega_{\bo\alpha},~\omega_{-\bo\alpha})$, 
$ \boldsymbol{\tau}_3=
\text{diag}(
 \mathbf{I}_{N\times N}, -\mathbf{I}_{N\times N} )$, and $\omega_{\bo\alpha}$ are the  energy eigenvalues and $\alpha$ labels the bands.
From Eq.~\ref{eqna} we get $\mathcal{P}_\bo^\dg= \boldsymbol{\tau}_3 \mathcal{P}_\bo^{-1} \boldsymbol{\tau}_3$, and Eq.~\ref{eqn1} is equivalent to saying that we need to diagonalize the Hamiltonian $\boldsymbol{\mathcal{H}}^\prime(\bo)= \boldsymbol{\tau}_3\boldsymbol{\mathcal{H}}(\bo),$ whose eigenvalues are given by $ \boldsymbol{\tau}_3\mathcal E_\bo$ and the columns of $\mathcal P_\bo$ are the corresponding eigenvectors. The eigenvalues of this Hamiltonian cannot be obtained analytically except at zero field.  The paraunitary operator  $\mathcal P_\bo$ defines a Berry curvature given by
 \begin{align}
 \Omega_{ij;\alpha}(\bo)=-2\text{Im}[ \boldsymbol{\tau}_3\mathcal (\partial_{k_i}\mathcal P_{\bo\alpha}^\dg) \boldsymbol{\tau}_3(\partial_{k_j}\mathcal P_{\bo\alpha})]_{\alpha\alpha},
 \label{bc1}
 \end{align}
with $i,j=\lbrace x,y\rbrace$ and $\mathcal P_{\bo\alpha}$ are the columns of $\mathcal P_{\bo}$. In this form, the Berry curvature simply extracts the diagonal components which are the most important.  From Eq.~\ref{eqn1} the Berry curvature can be written alternatively as

\begin{align}
\Omega_{ij;\alpha}(\bold k)=-\sum_{\alpha^\prime\neq \alpha}\frac{2\text{Im}[ \braket{\mathcal{P}_{\bo\alpha}|v_i|\mathcal{P}_{\bo\alpha^\prime}}\braket{\mathcal{P}_{\bo\alpha^\prime}|v_j|\mathcal{P}_{\bo\alpha}}]}{\lb\omega_{\bo\alpha}-\omega_{\bo\alpha^\prime}\rb^2},
\label{chern2}
\end{align}
where $\bold v=\partial\boldsymbol{\mathcal{H}}^\prime(\bo)/\partial \bold k$ defines the velocity operators. The Berry curvature is related to DM interaction $\Omega(\bold k)\propto \phi$ and the Chern number is defined as,
 \begin{equation}
\mathcal{C}_\alpha= \frac{1}{2\pi}\int_{{BZ}} dk_xdk_y~ \Omega_{xy;\alpha}(\bold k).
\label{chenn}
\end{equation}


\begin{thebibliography}{0}
 \bibitem{kru} V. V. Kruglyak, S. O. Demokritov, and D. Grundler, J.
Phys. D: Appl. Phys. {\bf 43}, 264001 (2010).
  
   \bibitem{kru2} 
  Benjamin Lenk, Henning Ulrichs, Fabian Garbs, Markus M\"{u}nzenberg, Physics Reports {\bf 507}, 107 (2011).
   \bibitem{kru1} 
A. V. Chumak {\it et al}., Nature Physics {\bf 11}, 453 (2015).
 \bibitem{dm}
 { I. Dzyaloshinsky}  {J. Phys. Chem. Solids} {\bf 4}, {241} {1958}; {T. Moriya} {Phys. Rev. }{\bf 120},  91 (1960).
  \bibitem{dm1}
 {T. Moriya} {Phys. Rev. }{\bf 120},  91 (1960).
 \bibitem{alex0}
{H. Katsura, N. Nagaosa, and P. A. Lee} {Phys. Rev. Lett.} {\bf 104},  {066403} {(2010)}.
\bibitem{alex1}
 Y. Onose {\it et al}.,  Science  { \bf 329}, 297 (2010).
\bibitem{alex1a}
T. Ideue {\it et al}., Phys. Rev. Lett. {\bf 85}, 134411 (2012).
  \bibitem{alex6}
 Max Hirschberger {\it et al}.,  \prl {\bf 115}, 106603 (2015).
  \bibitem{alex6a}
R. Chisnell {\it et al}., Phys. Rev. Lett. {\bf 115}, 147201 (2015).
 \bibitem{alex2}
  {R. Matsumoto and S. Murakami} {Phys. Rev. Lett.}  {\bf 106},  {197202} {(2011)}; {Phys. Rev. B} {\bf 84},  {184406} {(2011)}.
 \bibitem{alex7}
 A.  A. Kovalev and V.  Zyuzin, Phys. Rev. B {\bf 93}, 161106(R) (2016).


 
 \bibitem{hir}
M. Hirschberger {\it et al}.,  Science {\bf 348}, 106 (2015). 
  \bibitem{shin}
 R. Shindou et.al., Phys. Rev. B {\bf 87},  174427 (2013); Phys. Rev. B {\bf 87}, 174402 (2013).
  \bibitem{shin1}
R. Matsumoto, R. Shindou, and S. Murakami, Phys. Rev. B {\bf 89}, 054420 (2014).
  \bibitem{zhh} 
  L. Zhang {\it et al}., Phys. Rev. B {\bf 87}, 144101 (2013).
\bibitem{alex4a}
 { H. Lee, J. H. Han, and P. A. Lee}   {Phys. Rev. B.} {\bf 91}, 125413  {(2015)} .
\bibitem{alex4}
 {A.  Mook, J.  Henk, and I. Mertig} {Phys. Rev. B} {\bf 90}, 024412 (2014); A.  Mook, J.  Henk, and I. Mertig,  {Phys. Rev. B} {\bf 89}, 134409 (2014).
 \bibitem{jh}
 J. H. Han, H. Lee, 	J. Phys. Soc. Jpn. {\bf 86}, 011007 (2017).
 \bibitem{sol}
S. A.  Owerre, J. Phys.: Condens. Matter {\bf 28}, 386001  (2016).

\bibitem{sol1}
S. A.  Owerre, J. Appl. Phys. {\bf 120}, 043903 (2016).
\bibitem{kkim}
Se Kwon Kim {\it et al}., Phys. Rev. Lett. 117, 227201 (2016).

%
 
 \bibitem{mok}
 A. Mook, J. Henk, and I. Mertig, Phys. Rev. Lett. {\bf 117}, 157204 (2016).
 \bibitem{su}
 Ying Su, X. S. Wang, X. R. Wang, arXiv:1609.01500 (2016). 
  \bibitem{zheng1}
 H. Yao and S. A. Kivelson, Phys. Rev. Lett. {\bf 99}, 247203  (2007).
 \bibitem{zheng6}
Heng-Fu Lin {\it et al}., Phys. Rev. A {\bf 90}, 053627 (2014).
\bibitem{zheng}
Y. -Z. Zheng {\it et al}.,  Angew. Chem., Int. Ed. {\bf 46}, 6076 (2007).



   \bibitem{zheng3}
 J. O. Fjaerestad, arXiv:0811.3789.
   \bibitem{zheng4}
 B.-J. Yang, A. Paramekanti, and Y. B. Kim, Phys. Rev. B {\bf 81}, 134418 (2010).

  \bibitem{zheng2}
 A. R\"{u}egg, J. Wen, and G. A. Fiete, Phys. Rev. B {\bf 81}, 205115 (2010).
\bibitem{zheng5}
Wen-Chao Chen {\it et al}., Phys. Rev. B {\bf 86}, 085311 (2012).
\bibitem{zheng5a}
Mengsu Chen and Shaolong Wan, J. Phys.: Condens. Matter {\bf 24}, 325502 (2012).
  \bibitem{HP}
T. Holstein and H. Primakoff, Phys. Rev. {\bf 58}, 1098 (1940).
 \bibitem{fdm}
 {F. D. M. Haldane} {Phys. Rev. Lett.} {\bf 61}, {2015}  {(1988)}. 

  \bibitem{mag}
J. Fransson, A. M. Black-Schaffer, and A. V. Balatsky, Phys. Rev. B {\bf 94}, 075401 (2016).

\bibitem{luuk}
Luuk J. P. Ament, Michel van Veenendaal, Thomas P. Devereaux, John P. Hill, and Jeroen van den Brink, Rev. Mod. Phys. {\bf 83}, 705 (2011).
\bibitem{kha}
Khalil Zakeri, Physics Reports {\bf 545}, 47 (2014).


\end{thebibliography}
\end{document}